\documentstyle[spie]{article}

\title{Hamiltonians for Quantum Computing}

\author{Vladimir Privman\supit{a}, Dima Mozyrsky\supit{a}, and Steven P. Hotaling\supit{b}
\skiplinehalf
\supit{a}Department of Physics, Clarkson University, Potsdam, New York\hspace{0.5em}13699-5820
\skiplinehalf
\supit{b}Air Force Materiel Command, Rome
Laboratory/Photonics Division
\skipline
25 Electronic Parkway, Rome, New York\hspace{0.5em}13441-4515}

\begin{document}
\maketitle

\begin{abstract}
We argue that the analog nature of quantum computing makes the usual
design approach of constructing complicated logical operations from 
many simple gates inappropriate. Instead, we propose to design multi-spin 
quantum gates in which the input and output two-state systems (spins) 
are not necessarily identical. We outline the design criteria for such
devices and then review recent results for single-unit Hamiltonians 
that accomplish the NOT and XOR functions.
\end{abstract}

\keywords{Quantum computing, Analog computing, Hamiltonians for quantum gates}

\section{INTRODUCTION}

One of the great challenges of the physics of nanoscale systems
has been the design of atomic-size devices operating in a quantum-coherent fashion.
Dimensions of semiconductor computer components will soon reach\cite{1}
about $0.25\,\mu$m$\,=2500\,$\AA, which is well above
the atomic sizes at which quantum-mechanical effects are important.
However, it is generally expected that as the miniaturization continues,
atomic dimensions will be reached. This article concerns with quantum computing,
i.e., nanoscale devices that perform logical operations while maintaining
quantum coherence. Some early studies\cite{2,3,4,8,9} considered how
quantum mechanics affects the foundations of computer science;
issues such as limitations on classical computation due to
quantum fluctuations, etc., have been raised.
A more recent development\cite{4,5,6,7,8,9,10,11,12,13,14,15,16,17,18,19,20,21,22,23,24,25,26,27,28,29,30,30a,30b} 
has been to utilize the quantum nature of components of atomic
dimensions for more efficient computations involving quantum-coherent evolution.

Quantum computing has attracted a lot of interest recently
owing to several new features. Firstly, it may be faster than
classical computing: new fast quantum algorithms have
been proposed.\cite{31,32,33,34,35} Error
correction techniques,\cite{10,27,31,36,37,37a,37b,37c} unitary operations corresponding
to the simplest logic gates,\cite{5,6,7,8,9,10,11,12,13,14,15,16,17,18,19,20,21,22,23,24,25,26,27,28,29,30,30a,30b} and
some Hamiltonians for gate
operation\cite{10,11,14,24,28,29,30,30a,30b,37d} have
been explored. Ideas on how to combine the simplest quantum gates
have been put forth.\cite{7,15,38} Experimentally,
there are
several atomic-scale systems where the simplest quantum-gate functions
have been recently realized\cite{26,39,40} or contemplated.\cite{19}

There remain, however, many conceptual difficulties with
quantum computing.\cite{4,18} The reversibility of coherent quantum
evolution implies that the time scale $\Delta t$ of the operation of
quantum logic gates must be built into the Hamiltonian. As a result,
all the proposals available to date assume that computation will
be externally timed, i.e., interactions will be switched {\it on\/}
and {\it off}, for instance, by laser radiation.

This means that if logical
operations are constructed from one or few simple universal gates, then
each such gate will have to be precisely controlled from outside. In ordinary
classical (i.e., macroscopic, irreversible) computing, the NAND gate is
an example of a universal gate. From it complicated logical operations
can be constructed. In the classical case, however, it is the internal relaxation
processes in the basic gate(s) that determine the time scale of their
operation (equilibration) $\Delta t$. We consider it extremely unlikely
that one would ever be able to control externally, in a coordinated fashion,
millions of simple reversible quantum gates in order to operate a macroscopic
computer.

Furthermore, quantum computers are naturally {\it analog}\cite{22} in their
operation. Indeed, in order to use the power of quantum interference (superposition
of states), one has to allow any linear combination of the basis qubit
states $|1\rangle$ and $|0\rangle$. Analog errors 
are difficult to correct. By analog errors we mean those minor variations
in the input and output variables which cannot on their own be identified as erroneous
in an analog device because its operation involves {\it continuous\/} values of
variables (so that the fluctuated values are as legal as the original ones). 
By noise errors we term those that result from single-event
problems with device operation, or from external influences (including decoherence
in the quantum case), or from other failures in operation. All errors
in a {\it digital\/} device (i.e., deviations from discrete values) 
can be systematically decreased or eliminated
in each step of a calculation. Similarly, the noise errors in analog devices
can be corrected or decreased.

However, the analog errors cannot be corrected.
Consider a state $\alpha |1\rangle + \beta |0\rangle$
and a nearby state $\alpha^\prime |1\rangle + \beta^\prime 
|0\rangle+\sum_j \zeta_j |j\rangle$, where
$\alpha^\prime$ is close to $\alpha$, $\beta^\prime$ is close to $\beta$, while
$\zeta_j$ are small. The latter terms represent admixture of quantum states 
$|j\rangle$ other than the two qubit states. Both
states are equally legal as input and output quantum
states. We could restrict input or
output to a vicinity of certain states, for instance, the basis states
$|1\rangle$ and $|0\rangle$, thus moving towards digitalization. However,
we then loose the quantum-interference property. Another important effect: 
decoherence, that would require a density
matrix description, falls in the noise-error category.

Modern error-correction
techniques\cite{10,27,31,36,37,37a,37b,37c} can handle the noise
errors but not the analog errors. To illustrate, consider
this quote\cite{37c} from the article entitled
{\it Quantum Error Correction for Communication}:
``To achieve this the sender can add two qubits, initially both in state
$|0\rangle$, to the original qubit and then perform an encoding unitary
transformation$\ldots$''. The problem here is that the states actually
encountered in the system during error correction
are not available as basis qubit states  (such as $|0\rangle$) with 
infinite precision. Typically, by qubits we mean a set of two orthogonal
quantum states selected from the energy eigenstates of the system. Even assuming that
the thermal noise can be reduced at low temperatures to make the ground state sufficiently
long-lived, the excited states of any system, especially if it is a part of a 
macroscopic computer, will not be defined sharply enough to provide ideal
stationary states $|1\rangle$ and $|0\rangle$. External interactions, spontaneous
emissions, etc., will generate both noise- and analog-errors in the
basis states, i.e., the actual state (disregarding decoherence)
will be $\alpha |1\rangle + \beta 
|0\rangle+\sum_j \zeta_j |j\rangle$,
with $\alpha \simeq 1$ and $\beta , \zeta_j \simeq 0$, instead of the ideal 
$|1\rangle$ which is an eigenstate of an ideal, isolated-system Hamiltonian.

Furthermore, analog errors will be magnified
when separate simple-gate operations are combined to yield
a complex logical function. Thus, the conventional picture
of a quantum computer is unrealistic: it assumes a multitude
of simple-gate units each being externally controlled by laser beams
(one needs a lot of graduate students for that!). Such computers will 
magnify analog errors which cannot be corrected in principle because
the error state is as legal as the original state.

In this work we therefore adopt a 
view typical of the analog-computer approach, of designing the computer
as a single unit performing in one shot a complex logical task
instead of a chain of simple gate tasks. This approach will not repair
all the ailments outlined earlier. For instance, the computer as a whole will
still be subject to analog errors. However, these will not be magnified
by proliferation of sub-steps each of which must be exactly controlled.

In fact, we consider it likely that 
technological advances might first allow design and manufacturing of
limited-size units, based on several tens of atomic two-level systems,
operating in a quantum-coherent fashion over a sufficiently large time
interval to function as parts of a larger classical (dissipative) computer which
will not maintain a quantum-coherent operation over its
macroscopic dimensions. We would like these to function as single 
analog units rather then being 
composed of many gates. 

The outline of this review is as follows. In Section 2 we continue our 
discussion of the design of quantum gates. In Section 3 we review known
results for the simplest NOT gate mainly to set the notation and nomenclature.
A more complicated, two-spin NOT gate is studied in Section 4. Section 5
addresses  the time-dependence of the Hamiltonians. Finally, Sections
6, 7, 8 review results for a three-spin XOR gate.

\section{DESIGN CONSIDERATIONS FOR MULTI-SPIN QUANTUM GATES}

In order to make
connections with the classical computer-circuitry design
and identify, at least initially,
which multi-qubit systems are of interest,
we propose to consider spatially extended multi-spin quantum 
gates with input and output qubits possibly different. The reason for 
emphasizing this property is that multi-spin devices will have spatial
extent. The interactions that feed the input need not be identical to
those interactions/measurements that read off the output. Furthermore,
for systems with short-range interactions one can only access the boundary
spins in a large cluster. Thus we may use only part of the spins to specify
the input and another subset to contain the output. The two sets may be identical,
partially overlapping, or nonoverlapping.

Reversibility of coherent quantum evolution makes the
distinction between the input and output less important than in
irreversible
computer components. However, we consider the notion of separate (or
at least not necessarily fully identical) input and output useful within
our general goals: to learn what kind of interactions are involved
and to consider also units that might be connected to/as in
classical (dissipative) computer devices.

Our goal is to be able to design interaction parameters,
presumably by numerical simulations, to have such gates perform
useful Boolean operations. This is not an easy task. Actually,
it must be broken into several steps. First, we must identify those
interactions which can be realized in solid state or other
experimental arrangements. As examples below illustrate for several
simplest gates the form of the interaction Hamiltonians is quite 
unusual by the solid-state standards.

Secondly, we expect interactions to be short-range and two-particle
(two-spin) when several two-state systems (termed qubits, spins) are
involved. 

Thirdly, incorporating designed coherent
computational units in a larger classical computer will require a whole
new branch of computer engineering because the built-in Boolean
functions will be complicated as compared to the conventional
NOT, AND, OR, NAND, etc., to which computer designers
are accustomed. Furthermore, the rules of their
interconnection with each other and with the rest of the classical
computer will be different from today's devices.

Our initial studies have been analytical. In the future 
we foresee numerical studies of systems of order 20 to 25 two-state
(spin) atomic components with variable general-parameter interactions. 
In this review we summarize results\cite{28,30a,30b} for
interaction Hamiltonians required for
operation of the NOT (Sections 3,4) and XOR (Sections 6, 7, 8) 
logic gates. Other results presently available 
include Hamiltonians for certain NOT\cite{14,28} and 
controlled-NOT gates\cite{10,30,37d}, and for some
copying processes\cite{29,30}, as well as general analyses of possibility
of construction of quantum computing systems\cite{8,22}. 

Quantum logics and the dynamics of quantum gates should
be fully reversible. Implications of this property have
biased recent literature on the quantum logic gates. Firstly, the
distinction between the input and
the output parts of
the system has been blurred. A typical configuration involves a
quantum-mechanical system that is
``programmed'' with the input and
then after the time interval $\Delta t$ it will
be in the output state.
We note that the time interval $\Delta t$ is fully determined by the
parameters of the Hamiltonian; in
order to effect the quantum gate operation, the interaction energies
associated with both the internal and external-field parts of the
Hamiltonian must be of order $\hbar / \Delta t$.

Consideration of multi-spin quantum gates requires a large
number of basis states. However, it is also useful to study
few-spin exactly solvable systems. These provide explicit
examples of what the actual interaction Hamiltonians
should look like. A notable exactly solvable system, 
known before the quantum-computing
field became active, is the NOT gate operation 
in a two-state qubit\cite{14}
obtained by applying a constant external magnetic field
to a single spin. Then another field is applied, oscillating in time,
in a direction perpendicular to the constant field. This
{\it paramagnetic resonance\/} problem is a textbook
example of time-dependent quantum-mechanical evolution. 

An accepted approach has been to
consider interactions switched on only for the duration of the gate
operation $\Delta t$. If the gate is actually the whole
computer then one can regard the interaction as time-independent.
However, for specific tasks in components with a limited number
of basis states, it may be appropriate to view the interaction as
controlled externally to be switched on and off. While general ideas of
externally timed computation are not new,\cite{4} 
actual realizations in quantum computation with many sub-unit gates
will encounter difficulties outlined earlier. 
General developments for the latter type of interaction
(time-independent or on/off) have included\cite{8,22} identification
of unitary operators that correspond to quantum computer operation and
establishment of the existence of the appropriate interaction
Hamiltonians.

A quantum gate performs an
operation whereby the input state
determines the output
state after a time interval $\Delta t$. The interactions must
be controlled, i.e., switched on and off, in order 
to have the gate operation during the interval $\Delta t$
independent of the interactions with the computer parts external
to the gate. This control
of interaction, i.e., external timing of the computer operation
already mentioned earlier, can be possibly accomplished by
the external interactions while the internal interactions
be reserved for the gate operation. However, we would like to
consider multi-spin gates in order to avoid too many such
controlling external influences.

With regards to the
requirement to control the interactions externally, with the
time dependence given by the on/off protocol, we will show in Section 5
how to extend this approach to certain time-dependent
interactions (protocols) which are more smooth than the on/off
shape. 

\section{THE SIMPLEST NOT GATE}
 
In this section we consider the NOT gate\cite{28} based on a single qubit. 
This gate has been extensively studied in the 
literature.\cite{5,7,13,14,15,22}
Our presentation here is only intended to set up the
notation and illustrate methods useful in more complicated situations
addressed in Sections 4, 6, 7, 8.
We label by $\pmatrix{1\cr 0}$ and
$\pmatrix{0\cr 1}$ the two basis states.
The NOT gate corresponds to those
interactions which, over the time interval
$\Delta t$, accomplish the following:

$$ \pmatrix{1\cr 0} \Longrightarrow
e^{i\alpha}\pmatrix{0\cr 1} \, ,\eqno(1) $$

$$ \pmatrix{0\cr 1} \Longrightarrow
e^{i\beta}\pmatrix{1\cr 0} \, .\eqno(2) $$

\noindent Here $\alpha$ and $\beta$ are
arbitrary. The unitary matrix $U$
that corresponds to this evolution is

$$ U=\pmatrix{0&e^{i\beta}\cr
 e^{i\alpha}&0} \, , \eqno(3) $$

\noindent with the eigenvalues

$$ u_1=e^{i(\alpha+\beta)/2} \qquad {\rm and}
\qquad u_2=-e^{i(\alpha+\beta)/2} \, , \eqno(4)$$

\noindent while the normalized eigenvectors yield
the transformation matrix $T$ which
can be used to diagonalize $U$:

$$ T={1\over
\sqrt{2}}\pmatrix{e^{i\beta/2}&e^{i\beta/2}\cr
 e^{i\alpha/2}&-e^{i\alpha/2}} \, . \eqno(5)$$

\noindent We have 

$$ T^\dagger U T = \pmatrix{u_1&0\cr
 0&u_2} \, . \eqno(6) $$

\noindent Here the dagger superscript denotes Hermitian conjugation.

We next use the relation

$$ U=e^{-iH\Delta t/\hbar}  \eqno(7) $$

\noindent for a time-independent
Hamiltonian. In the diagonal representation, it 
yields the energy levels:

$$ E_1=-{\hbar \over 2 \Delta t}(\alpha+ \beta)+{2\pi \hbar
\over \Delta t}N_1 \, , \;\; \;\;\;
 E_2=-{\hbar \over 2 \Delta t}(\alpha+ \beta)+{2\pi \hbar
\over \Delta
t}\left(N_2+{1\over 2}\right) \, , \eqno(8)$$

\noindent where $N_1$ and $N_2$ are
arbitrary integers. The Hamiltonian is then obtained
from the relation

$$H=T\pmatrix{E_1&0\cr
 0&E_2}T^\dagger  \eqno(9) $$

\noindent as a certain $2\times 2$ matrix. The latter is
conveniently represented is terms of
the unit matrix $\cal I$ and the conventional Pauli
matrices $\sigma_x$, $\sigma_y$, $\sigma_z$:

$$ H=\left[-{\hbar \over 2 \Delta t}(\alpha+ \beta)+{\pi
\hbar\over\Delta t}\left(N_1+N_2+{1\over 2}\right)
\right]{\cal I}
+{\pi \hbar\over\Delta t}\left(N_1-N_2-{1\over 2}\right)\left[
\left(\cos{\alpha-\beta\over
2}\right)\sigma_x+\left(\sin{\alpha-\beta\over 2}\right)\sigma_y\right]
\, . \eqno(10)$$

To effect the gate operation, the interaction must be switched on
for the time interval $\Delta t$. The constant part of the Hamiltonian
only affects the average phase $\alpha+\beta \over 2$
of the transformation
(1)-(2). Thus this term can be disregarded.

The nontrivial part of (10) depends on the integer $N=N_1-N_2$ which
is arbitrary, and on one arbitrary variable

$$ \gamma = {\alpha- \beta\over 2} \, . \eqno(11) $$

\noindent Thus we can use the Hamiltonian

$$ H={\pi \hbar\over\Delta t}\left(N-{1\over 2}\right)\left[
\left(\cos{\gamma}\right)\sigma_x
+\left(\sin{\gamma}\right)\sigma_y\right] \, .\eqno(12) $$

\noindent For a spin-$1\over 2$ two-state
system such an interaction can be obtained
by applying a magnetic field oriented
in the $xy$-plane at an angle $\gamma$ with the $x$-axis.
The strength of the field is inversely proportional to the desired
time interval $\Delta t$, and
various allowed field values are determined by
the choice of $N$.

We note that during application of the external field the {\it up\/}
and {\it down\/}
quantum states in (1)-(2) are not the
eigenstates of the Hamiltonian.
If the time interval $\Delta t$ is not short enough, the energy-level
splitting $|E_1-E_2|\propto
|N-{1\over 2}|$ can result in spontaneous emission
which is one of the sources of errors in computer
operation.
Generally, when implemented in condensed
matter, the
two states of the qubit may be part of a spectrum of many energy levels.
In order to minimize the number of
spontaneous transition modes,
the best choice of the interaction strength would
correspond to minimizing
$|E_1-E_2|$, i.e., to $|N-{1\over 2}|={1\over 2}$.
 
\section{THE SPATIALLY EXTENDED TWO-SPIN NOT GATE}
 
In this section we consider a more complicated situation. Two
two-state systems,
input ($I$) and output ($O$), are involved. We will use the
following self-explanatory notation for the state vector:

$$ \pmatrix{a_1\cr a_2\cr a_3\cr a_4}
=a_1 |\uparrow\uparrow\rangle+
a_2 |\uparrow\downarrow\rangle+
a_3 |\downarrow\uparrow\rangle +a_4 |\downarrow\downarrow\rangle 
\qquad\qquad\qquad\qquad\qquad\qquad\qquad\qquad\qquad\qquad\qquad\qquad\qquad\qquad $$ $$
\qquad\qquad\qquad\qquad=a_1\pmatrix{1\cr 0}_I \otimes \pmatrix{1\cr 0}_O +
a_2\pmatrix{1\cr 0}_I \otimes \pmatrix{0\cr 1}_O
+a_3 \pmatrix{0\cr 1}_I \otimes \pmatrix{1\cr 0}_O
+a_4 \pmatrix{0\cr 1}_I \otimes \pmatrix{0\cr 1}_O \, . \eqno(13) $$

\noindent In what follows we will
omit the direct-product symbols $\otimes$ when multiplying expressions
with subscripts $I$ and $O$.

We seek transformations such that irrespective of the initial state of
$O$, and provided $I$ is initially in the up or down state,
the final state has $O$ in the down or up state, respectively (while
the final state of $I$ is not restricted). Thus, we place the logical
NOT of $I$ in $O$ provided the initial state of $I$ was
one of the basis states corresponding to the classical bit values 1 and 0.
The desired transformation maps any state with $a_3=a_4=0$ into
a state with components 1 and 3 equal zero, i.e., input up
yields output down. Similarly, any state with $a_1=a_2=0$
should evolve into a state with components 2 and 4 equal zero,
corresponding to input down giving output up.
The general evolution operator must therefore be of the form

$$U=\pmatrix{0&0&U_{13}&U_{14}\cr
 U_{21}&U_{22}&0&0\cr
 0&0&U_{33}&U_{34}\cr
 U_{41}&U_{42}&0&0} \, , \eqno(14) $$

\noindent which depends on 16 real parameters. However, one can show that the
requirement of unitarity, $U^\dagger U=1$,
imposes 8 conditions so that the number of
real parameters is reduced to 8. The following parametrization
covers all such unitary matrices:

$$U=\pmatrix{0&0&e^{i\chi}\sin\Omega&e^{i\beta}\cos\Omega\cr
-e^{i(\alpha+\rho-\eta)}\sin\Upsilon&e^{i\rho}\cos\Upsilon&0&0\cr
0&0&e^{i\delta}\cos\Omega&-e^{i(\beta+\delta-\chi)}\sin\Omega\cr
e^{i\alpha}\cos\Upsilon&e^{i\eta}\sin\Upsilon&0&0} \, . \eqno(15) $$

\noindent Here all the variables are unrestricted; we could
limit $\Omega$ and $\Upsilon$ to the range $\left[0,{\pi \over
2}\right]$ without loss of generality.

In order to make the calculation analytically tractable, we will
restrict the number of free parameters to four by considering the matrix

$$U=\pmatrix{0&0&0&e^{i\beta}\cr
 0&e^{i\rho}&0&0\cr
 0&0&e^{i\delta}&0\cr
 e^{i\alpha}&0&0&0} \, . \eqno(16) $$

\noindent This form has been favored for a possible analytical calculation for 
the following reasons. Firstly, the
structure of a single phase-factor in each column is similar to that of
the two-dimensional (single-spin) matrix encountered earlier. Secondly, the form
(16) contains Hermitian-$U$ cases ($\beta=-\alpha$, $\rho=0$ or $\pi$,
$\delta=0$ or $\pi$). Therefore, the eigenvalues, which are generally
on the unit circle for any unitary matrix, may be positioned
symmetrically with respect to the real axis, as functions of the
parameters. Indeed, the eigenvalues of $U$ turn out to be quite simple:
 
$$ u_1=e^{i(\alpha+\beta)/2}\, , \;\; \;\;
u_2=-e^{i(\alpha+\beta)/2} \, , \;\; \;\;
u_3=e^{i\rho} \, , \;\; \;\;
u_4=e^{i\delta} \, . \eqno(17) $$

\noindent The (unitary) diagonalizing matrix $T$ is

$$ T={1\over
\sqrt{2}}\pmatrix{e^{i\beta/2}&e^{i\beta/2}&0&0\cr
 0&0&\sqrt{2}&0\cr
 0&0&0&\sqrt{2}\cr
 e^{i\alpha/2}&-e^{i\alpha/2}&0&0} \, . \eqno(18)$$

The next step is to identify the energy levels.
We chose the notation such that the energies $E_{1,2}$ are identical
to (8). The other two energies are given by

$$ E_3=-{\hbar \over \Delta t}\rho +{2\pi \hbar
\over \Delta t}N_3 \, , \;\; \;\;
E_4=-{\hbar \over \Delta t}\delta +{2\pi \hbar
\over \Delta t}N_4 \, . \eqno(19)$$

\noindent The Hamiltonian is then obtained
as in the single-spin NOT case. It is convenient to
avoid cumbersome expressions by expressing it in terms of the energies;
the latter will be replaced by explicit expressions (8), (19) when
needed. The resulting $4\times 4$ matrix has been expressed in terms of
the direct products involving the unit matrices and the Pauli matrices
of the input and output two-state systems. 
We only report the result:

$$ H={1\over 4}\left(2E_1+2E_2+E_3+E_4\right)
+{1\over
4}\left(E_3-E_4\right)\left(\sigma_{zI}-\sigma_{zO}\right)
+{1\over
4}\left(2E_1+2E_2-E_3-E_4\right)\sigma_{zI}\sigma_{zO} 
\quad\qquad\quad\qquad$$ $$\quad\qquad\quad\qquad
+{1\over 4}\left(E_1-E_2\right)\left(\cos{\alpha-\beta\over 2}\right)
\left(\sigma_{xI}\sigma_{xO}-\sigma_{yI}\sigma_{yO}\right)
+{1\over 4}\left(E_1-E_2\right)\left(\sin{\alpha-\beta\over 2}\right)
\left(\sigma_{xI}\sigma_{yO}+\sigma_{yI}\sigma_{xO}\right)
\, .\eqno(20) $$

\noindent The constant part of the Hamiltonian
can be changed independently of the other coupling constants and it can
be discarded. We can also generally vary the integers
$N_{1,2,3,4}$ and the variables $\alpha$, $\beta$, $\rho$, $\delta$.
The constant part is in fact proportional to ${\cal I}_I \otimes
{\cal I}_O$. We avoid this notation and present the
terms in the Hamiltonian in a more physically transparent form.

The Hamiltonian in (20) has also terms linear in the Pauli matrices (in
the spin components for spin systems). These correspond to interactions
with externally applied fields which in fact must be of opposite
direction for the $I$ and $O$ spins. As explained
in the
introduction, we try to avoid such interactions: hopefully, external
fields will only be used for clocking of the computation, i.e.,
for controlling the internal interactions via some intermediary part
of the system connecting the $I$ and
$O$ two-state systems. Thus, we will assume that $E_3=E_4$ so that there are no
terms linear is the spin components.

Among the remaining interaction terms, the term involving the
$z$-components in the product form $\sigma_{zI}\sigma_{zO}$ ($\equiv
\sigma_{zI}\otimes\sigma_{zO}$), has an arbitrary coefficient to be denoted
$-\cal E$. The terms of order two in the $x$ and $y$ components have
free parameters similar to those in (11)-(12).
The final expression is 

$$H=-{\cal E}\sigma_{zI}\sigma_{zO}+
{\pi \hbar\over 2\Delta t}\left(N-{1\over 2}\right)\Big[
\left(\cos{\gamma}\right)\left(\sigma_{xI}
\sigma_{xO}-\sigma_{yI}\sigma_{yO}\right)+\left(\sin{\gamma}\right)
\left(\sigma_{xI}\sigma_{yO}+\sigma_{yI}\sigma_{xO}\right)\Big]
\, .\eqno(21) $$

\noindent Here $N=N_1-N_2$ must be an integer. In order to
minimize the spread of the energies
$E_1$ and $E_2$ we could choose $|N-{1\over 2}|={1\over 2}$. 
Recall that we already have $E_3=E_4$. Thus
the energy levels of the Hamiltonian in (21) are

$$ E_1=-{\cal E}+{\pi \hbar\over \Delta t}\left(N-{1\over 2}\right)
\, , \;\; \;\; E_2=-{\cal E}-{\pi
\hbar\over \Delta t}\left(N-{1\over 2}\right)
\, , \;\; \;\; E_{3,4}={\cal E}\, . \eqno(22) $$

\noindent Thus degeneracy of three levels (but not all four) can be
achieved by varying the parameters.

The form of the interactions (21) is quite unusual as compared to
the traditional spin-spin interactions in condensed matter
models. The latter usually are based on the uniaxial (Ising)
interaction proportional to $\sigma_{z}
\sigma_z$, or the planar $XY$-model interaction proportional to
$\sigma_x\sigma_x +\sigma_y\sigma_y$, or the isotropic
(scalar-product)
Heisenberg interaction. The spin components here are those of two
different spins (not marked). The interaction (21) involves an
unusually high degree of anisotropy in the system. The $x$ and $y$
components are coupled in a tensor form which presumably will have to
be realized in a medium with well-defined directionality, possibly, a
crystal. 

\section{COMMENT ON TIME-DEPENDENCE OF INTERACTIONS}
 
The Hamiltonians considered thus far were all constant
for the duration of the gate operation. We note that the external control
of the interaction need not be limited to the time-dependence 
which is an abrupt on/off
switching. Indeed, we can modify the
time dependence according to

$$ H(t)=f(t)H \, , \eqno(23) $$

\noindent where we use the same symbol $H$ for both the original
time-independent interaction Hamiltonian such as (21) and the new,
time-dependent one, $H(t)$. The latter involves the protocol
function $f(t)$. The shape of this function, nonzero during the
operation of the
gate from time $t$ to time $t+\Delta t$, can be smooth.

For Hamiltonians involving externally applied fields, such as (12),
it may be important to have a constant plus an oscillatory components
(corresponding to constant and electromagnetic-wave magnetic fields,
for instance). However, 
the protocol function must satisfy

$$ \int\limits_t^{t+\Delta t}f(t')\, dt'=\Delta t \, , \eqno(24) $$

\noindent and therefore it cannot be purely oscillatory; it must have a
constant or other contribution to integrate to a nonzero value in
accordance with (24).

The possibility of the modification (23) follows from the fact that
the general relation

$$U=\left[e^{-i\int_t^{t+\Delta t}H(t')\, dt'/\hbar}
\right]_{\hbox{time-ordered}} \, \eqno(25)$$

\noindent does not actually require time ordering as long as the Hamiltonian
commutes with itself at different times. This condition is satisfied by
(23). Furthermore, if the Hamiltonian can be written as a sum of
commuting terms then each term can be multiplied by its own protocol
function. Interestingly, the Hamiltonian of the 
paramagnetic-resonance NOT gate\cite{14} is not of
this form. It contains a constant part and an oscillatory part but
they do not commute. Note that the term proportional to $\cal E$
in (21) commutes with the rest of that Hamiltonian. The terms proportional
to $\cos \gamma$ and $\sin \gamma$ do not commute with each other.
Rather, they anticommute, in (21), as such terms do in (12).

\section{THE THREE-SPIN XOR GATE}

Thus far we learned that extending the number of spins (qubits, two-state
systems) involved in the NOT gate from one to two produced an interaction
Hamiltonian family (21) with structure that is quite new 
and unfortunately not symmetric in terms of what we are used to in
solid-state magnetic interactions. We will now consider a {\it three-spin system}:
a quantum-XOR gate (which can also be realized\cite{10,30,37d} with
two spins). This choice is dictated by the fact that we can obtain analytical
results and address a new issue that was not there for one- or two-spin
systems: whether this quantum gate function can be accomplished with 
two-spin interactions.

We note that if a quantum logic operation is allowed to be decomposed
into a sequence of unlimited number of universal one- and two-spin
gates then one can always reduce it to two-spin interactions.\cite{5,7,15,38}
Here, however, we are interested in one-shot gates for which
the external control involves the overall system Hamiltonian, over a single
time interval $\Delta t$. The possibility of using solely two-spin interactions 
will actually depend on the logical function and for more complicated systems
it has to be explored by numerical studies. We note also that the issue of
having the interactions short-range (e.g., nearest-neighbor) does not really
arise for few-spin systems although it will be an important design
criterion as the number of spins (qubits) involved increases. Short-range
two-particle interactions are much better studied and accessible to
experimental probe than multi-particle interactions. 

We denote by $A$, $B$, $C$ the
three two-state systems, i.e., three spins (qubits). The transformation
must be specified for those initial states of  the input spins $A$ and $B$,
at time $t$,  that are one of the basis states $|AB\rangle=
|11\rangle$, $|10\rangle$, $|01\rangle$, or $|00\rangle$, where 1 and
0 denote the eigenstates of the $z$-components of the spin operators.
Here 1 refers to the up state and 0 refers to the down
state; we use this notation for consistency with the classical bit
notion. The initial state of $C$ is not specified. We would like to 
have a quantum evolution that mimics the XOR function:

$$\matrix{A&B&{\rm output}\cr{}1&1&0\cr{}1&0&1\cr{}0&1&1\cr{}0&0&0}
\eqno(26) $$

\noindent Here the output is at time $t+\Delta t$. One way to accomplish
this is to produce the output in $A$ or $B$, i.e.,
work with a two-spin system where the input and output are the same.
The Hamiltonian for such a system is not unique. Explicit examples
can be found\cite{10,30} where XOR was obtained as a sub-result of the
controlled-NOT
gate operation. In the case of two spins involved, the interactions
can be single- and two-spin only.

Here we require
that the XOR result be put in $C$ at time
$t+\Delta t$. The final states of $A$ and $B$, as well
as the phase of $C$ are arbitrary. In
fact, there are many different unitary transformations, $U$,
that correspond to the desired evolution in the eight-state space with
the basis $|ABC\rangle=|111\rangle$, $|110\rangle$, $|101\rangle$,
$|100\rangle$, $|011\rangle$, $|010\rangle$, $|001\rangle$,
$|000\rangle$. The choice of the
transformation determines what happens when the initial state is a
superposition of the reference states, what are the phases in
the output, etc.

Let us consider first the following Hamiltonian\cite{30a,30b}

$$ H={\pi \hbar \over 4 \Delta t}
 \left( \sqrt{2} \sigma_{zA} \sigma_{yB}
+ \sqrt{2} \sigma_{zB} \sigma_{yC}-\sigma_{yB} \sigma_{xC}
\right) \, . \eqno(27)$$

\noindent It is written here in terms of the spin components; the
subscripts $A,B,C$ denote the spins. In the
eight-state basis specified earlier, its matrix can be obtained by
direct product of the Pauli matrices and unit $2\times 2$ matrices
$\cal I$. For
instance, the first interaction term is proportional to 
$ \sigma_{zA}\otimes\sigma_{yB}\otimes{\cal I}_C$,
This Hamiltonian involves only two-spin-component interactions.
In fact, in this particular example $A$ and $C$ only interact with $B$.

One can show that the Hamiltonian (27) corresponds to the XOR result
in $C$ at $t+\Delta t$ provided $A$ and $B$ were in one of the
allowed superpositions of the appropriate binary states at $t$
(we refer to superposition here because $C$ is arbitrary at $t$).
There are two ways to verify this.\cite{30a,30b} Firstly, one can diagonalize
$H$ and then calculate the evolution matrix U in the diagonal 
representation by using
the general relation (7), valid for Hamiltonians which are constant
during the time interval $\Delta t$,
and then reverse the diagonalizing transformation.

The second, more general approach presented here is to
design a whole family of two-spin-interaction
Hamiltonians of which the form (27) is but a special case, by
analyzing generally a family of $8\times 8$ unitary
matrices corresponding to the three-spin XOR gate. 
This program is carried out in Sections 7, 8.

\section{THE STRUCTURE OF THE XOR UNITARY MATRIX AND HAMILTONIAN}

We require any linear combination of
the states $|\underline{11}1\rangle$ and
$|\underline{11}0\rangle$ to evolve
into a linear combination of $|11\underline{0}\rangle$,
$|10\underline{0}\rangle$, $|01\underline{0}\rangle$, and
$|00\underline{0}\rangle$; compare the underlined quantum numbers with
the first entry in (26), with similar rules for the other three entries
in (26). In the matrix notation, and in the standard basis
$|ABC\rangle=|111\rangle$, $|110\rangle$, $|101\rangle$,
$|100\rangle$, $|011\rangle$, $|010\rangle$, $|001\rangle$,
$|000\rangle$, the most general XOR evolution operator corresponding
to the Boolean function (26), with the output in $C$, is, therefore,

$$ U=\pmatrix{0&0&U_{13}&U_{14}&U_{15}&U_{16}&0&0\cr
 U_{21}&U_{22}&0&0&0&0&U_{27}&U_{28}\cr
 0&0&U_{33}&U_{34}&U_{35}&U_{36}&0&0\cr
 U_{41}&U_{42}&0&0&0&0&U_{47}&U_{48}\cr
 0&0&U_{53}&U_{54}&U_{55}&U_{56}&0&0\cr
 U_{61}&U_{62}&0&0&0&0&U_{67}&U_{68}\cr
 0&0&U_{73}&U_{74}&U_{75}&U_{76}&0&0\cr
 U_{81}&U_{82}&0&0&0&0&U_{87}&U_{88}} \, . \eqno(28)$$

\noindent The condition of unitarity, $UU^\dagger=1$, reduces the number of
independent parameters but they are still too numerous
for the problem to be manageable analytically; we are going
to consider a subset of operators of this form.

From our earlier discussion we know that one way to reduce
the number of parameters and ensure 
unitarity is to keep a single phase factor in each column and row of
the matrix. Some amount of lucky guessing is involved in finding
an analytically tractable parametrization. Thus, 
we choose a form which is diagonal in the
states of the $A$-spin,
$$ U=\pmatrix{V_{4\times 4}&0_{4\times 4}\cr
0_{4\times 4}&W_{4\times 4}} \, . \eqno(29) $$

\noindent Note that the input spins $A$ and $B$ are not treated symmetrically.
Here $0_{4\times 4}$ denotes the $4\times 4$ matrix of zeros.
The $4\times 4$ matrices $V$ and $W$ are parametrized as follows:

$$ V=\pmatrix{0&0&e^{i\delta}&0\cr
 e^{i\alpha}&0&0&0\cr
 0&0&0&e^{i\beta}\cr
 0&e^{i\gamma}&0&0} \, , \eqno(30) $$

$$ W=\pmatrix{0&e^{i\rho}&0&0\cr
 0&0&0&e^{i\omega}\cr
 e^{i\xi}&0&0&0\cr
 0&0&e^{i\eta}&0} \, . \eqno(31) $$

This choice of an 8-parameter unitary matrix $U$, 
see (29), was made because it has the structure

$$ 2U=\left(1+\sigma_{zA}\right)V+
 \left(1-\sigma_{zA}\right)W=
V+W+\sigma_{zA}(V-W) \, , \eqno(32)$$

\noindent where $V$ and $W$ are operators in the space of $B$ and $C$. Since
$U$ was chosen diagonal in the space of $A$, the Hamiltonian, $H$,
will have a similar structure,

$$ 2H=P+Q+\sigma_{zA}(P-Q) \, , \eqno(33)$$

\noindent with the appropriate Hamiltonians $P$ and
$Q$ in the $(B\otimes C)$-space. In order to avoid three-spin 
interactions, $P-Q$ must be linear in the
Pauli matrices. We also try to avoid single-spin
(external-field) interactions. Thus, $P+Q$ must contain only terms of
the second order in the spin components while $P-Q$ must contain only
terms of the first order in the spin components.
This suggests avoiding putting phase factors on the diagonal, which
would lead to matrices similar to those encountered in
extended-NOT-gate calculations that are known to be of a
structure undesirable here: they contain a mixture of first-order and
second-order terms. The off-diagonal choices remaining are
considerably limited: the forms
(30)-(31) are nearly unique. 

In summary, while the arguments are admittedly vague and they do
involve a certain level of guess, trial and error, the presented
parametrization offers a good chance that with further restrictions on
the parameters a two-spin interaction Hamiltonian can be obtained. As
will be seen shortly, five conditions are imposed so that the resulting
Hamiltonian depends on three (real) parameters.

Let us define

$$ \mu={\alpha+\beta+\gamma+\delta \over 4} \, , \;\; \;\; \nu={\rho+\omega+\xi+\eta \over 4} \, ,\eqno(34) $$

\noindent and introduce the reduced operators $p$ and $q$,

$$ P=-{\hbar \over \Delta t} p \quad \;\; {\rm and}
\quad \;\; Q=-{\hbar \over \Delta t} q \, , \eqno(35) $$

\noindent Then (7) yields

$$ V=\exp (ip) \quad \;\; {\rm and} \quad \;\; W=\exp(iq) \, . \eqno(36) $$

Next, we diagonalize $V$ and $W$: we calculate their eigenvalues and also
the matrices of their normalized eigenvectors, in order
to transform to the diagonal representations. 

Specifically, the eigenvalues of
$V$ are $e^{i\mu}$, $ie^{i\mu}$, $-e^{i\mu}$, $-ie^{i\mu}$. The
appropriate eigenvalues of $p$ then follow form (36) as $\mu$,
$\mu+{1\over 2}\pi$,
$\mu+\pi$, $\mu+{3\over 2}\pi$. Arbitrary multiples of $2\pi$ can
be added to these choices. However, there are certain nonrigorous
arguments for generally keeping the spread
of eigenvalues of the Hamiltonian as small as possible. Thus,
we choose the simplest expressions.
The eigenvalues of $q$ are determined identically, with $\mu$ replaced
by $\nu$ throughout.

The next step is to apply the inverse of the diagonalizing
transformations for $V$ and $W$ to the diagonal $4\times 4$ matrices
for, respectively, $p$ and $q$. The latter contain the eigenvalues of
$p$ and $q$ as the diagonal elements. The results are the matrix forms
of the operators $p$ and $q$ in the original representation: 

$$ {4 \over \pi}p=\pmatrix{ {4 \over \pi}\mu+3 &-(1+i)
 e^{i(\mu-\alpha)}&-(1-i) e^{i(\delta-\mu)}&- 
e^{i(2\mu-\alpha-\gamma)}\cr
-(1-i)e^{i(\alpha-\mu)}& {4 \over \pi}\mu+3 &- 
e^{i(2\mu-\beta-\gamma)}&-(1+i) e^{i(\mu-\gamma)}\cr
-(1+i) e^{i(\mu-\delta)}&-e^{i(\beta+\gamma-2\mu)}
& {4 \over \pi}\mu+3 &-(1-i) e^{i(\beta-\mu)}\cr
-e^{i(\alpha+\gamma-2\mu)}&-(1-i) e^{i(\gamma-\mu)}
&-(1+i) e^{i(\mu-\beta)}& {4 \over \pi}\mu+3 } \, , \eqno(37)$$

\ 

$$ {4 \over \pi}q=\pmatrix{ {4 \over \pi}\nu+3
&-(1-i) e^{i(\rho-\nu)}&-(1+i) e^{i(\nu-\xi)}
&- e^{i(\rho+\omega-2\nu)}\cr
-(1+i) e^{i(\nu-\rho)}& {4 \over \pi}\nu+3
&- e^{i(\omega+\eta-2\nu)}&-(1-i) e^{i(\omega-\nu)}\cr
-(1-i) e^{i(\xi-\nu)}&- e^{i(2\nu-\omega-\eta)}
& {4 \over \pi}\nu+3 &-(1+i) e^{i(\nu-\eta)}\cr
- e^{i(2\nu-\rho-\omega)}&-(1+i) e^{i(\nu-\omega)}&
-(1-i) e^{i(\eta-\nu)}& {4 \over \pi}\nu+3 } \, . \eqno(38)$$

\section{THE TWO-SPIN-INTERACTION XOR HAMILTONIAN}

Thus far we decreased the number of independent
parameters in the general unitary transformation and chose it to
be diagonal in the $A$-space. We now refine our design
of the Hamiltonian to favor interaction of the second
order in the Pauli matrices. First, we note that
both $P$ and $Q$ are constant-diagonal matrices. Therefore,
in terms of
the Pauli matrices both their sum and difference
in (33) will involve constant terms. These are undesirable
because in $\sigma_{zA}(P-Q)$ they lead to terms of order one (instead
of the desired two), in $H$, while in $P+Q$ they lead to an additive
constant in $H$ which only affects the overall phase of the
unitary transformation
and is of no interest otherwise. Therefore, we put 

$$ \mu=\nu=-{3\over 4}\pi \, , \eqno(39) $$

\noindent in order to nullify these diagonal elements in both $P$ and $Q$.

Let us now focus our attention on $P-Q$ which, by (39), is now
a matrix with zero diagonal. We must impose conditions on the
parameters to make
$P-Q$ of order exactly one in the Pauli matrices. We note, however,
that due to zero-diagonal, it cannot contain $\sigma_z$ terms. The
general form liner in $\sigma_x, \sigma_y$ is 

$$ P-Q={\cal I}_B \otimes \pmatrix{0&X\cr X^*&0}_C+
\pmatrix{0&Y\cr Y^*&0}_B\otimes{\cal I}_C =\pmatrix{0&X&Y&0\cr
X^*&0&0&Y\cr Y^*&0&0&X\cr 0&Y^*&X^*&0} \, , \eqno(40)$$

\noindent where the stars denote complex conjugation, $X$ and $Y$ are
arbitrary (complex) numbers, and $\cal I$ stands for the unit matrix
as before. Thus we require that $P-Q$ be of the form suggested
by (40). This imposes several 
conditions: two above-diagonal elements
of the difference must be equal to zero while the remaining four
elements must be equal pairwise. One can show that these conditions
are satisfied provided $\alpha, \beta, \gamma$ are kept as three
independent (real) parameters while the remaining angles are given by

$$ \delta=-3\pi-\alpha-\beta-\gamma \, , \eqno(41) $$
$$ \rho=-\pi+\beta \, , \eqno(42) $$
$$ \omega=-2\pi-\alpha-\beta-\gamma \, , \eqno(43) $$
$$ \xi=-\pi+\gamma \, , \eqno(44) $$
$$ \eta=\pi+\alpha \, . \eqno(45) $$

\noindent Note that (39) is built into (41)-(45). 
Given this choice, it turns out that
$P+Q$ contains only two-spin interactions. We have no simple explanation of this
property (of the absence of first-order terms in $P+Q$). It is
probably related to the fact that the structure pattern of the
original matrices $V$ and $W$ is quite similar even though the
precise positioning of nonzero elements is different.
We find

$$P+Q= - {\sqrt{2}\pi \hbar i \over 4 \Delta t} 
\pmatrix{0
&e^{-i\alpha}+e^{i\beta}
&e^{-i(\alpha+\beta+\gamma)}-e^{-i\gamma}
&-\sqrt{2}e^{-i(\alpha +\gamma)}\cr
-e^{i\alpha}-e^{-i\beta}
&0
&-\sqrt{2}e^{-i(\beta+\gamma)}
&e^{-i\gamma}-e^{-i(\alpha+\beta+\gamma)}\cr
e^{i\gamma}-e^{i(\alpha+\beta+\gamma)}
&\sqrt{2}e^{i(\beta+\gamma)}
&0
&-e^{-i\alpha}- e^{i\beta}\cr
\sqrt{2}e^{i(\alpha +\gamma)}
&e^{i(\alpha+\beta+\gamma)}-e^{i\gamma}
&e^{i\alpha}+e^{-i\beta}
&0} \, , \eqno(46) $$

$$ P-Q= - {\sqrt{2} \pi \hbar i \over 4 \Delta t} 
\pmatrix{0
&e^{-i\alpha}-e^{i\beta}
&e^{-i(\alpha+\beta+\gamma)}+e^{-i\gamma}
&0\cr
-e^{i\alpha}+ e^{-i\beta}&0&0&
e^{-i(\alpha+\beta+\gamma)}+e^{-i\gamma}\cr
-e^{i(\alpha+\beta+\gamma)}-e^{i\gamma}&0&0&e^{-i\alpha}-
e^{i\beta}\cr 0&-e^{i(\alpha+\beta+\gamma)}-e^{i\gamma}
&-e^{i\alpha}+ e^{-i\beta}&0} \, , \eqno(47) $$

Finally, we expand these matrices in terms of products of the Pauli
matrices and collect terms according to (33) to get

$$ H=-{\pi \hbar \over 8 \Delta t} \Big\{ \sqrt{2}
\left( \sin\alpha + \sin\beta \right) \sigma_{zA}\sigma_{xC} - \sqrt{2}
\left( \cos\alpha - \cos\beta \right) \sigma_{zA}\sigma_{yC} 
+ \sqrt{2}\big[ \sin\gamma + \sin(\alpha+\beta+\gamma)
\big] \sigma_{zA}\sigma_{xB} 
\qquad\qquad$$ $$\; - \sqrt{2}\big[ \cos\gamma +
\cos(\alpha+\beta+\gamma) \big] \sigma_{zA}\sigma_{yB}
+\sqrt{2}
\left( \sin\alpha - \sin\beta \right) \sigma_{zB}\sigma_{xC} - \sqrt{2}
\left( \cos\alpha + \cos\beta \right) \sigma_{zB}\sigma_{yC} 
\qquad\qquad\qquad\qquad\quad$$ $$\;\, - \sqrt{2}\big[ \sin\gamma-\sin(\alpha+\beta+\gamma) \big]
\sigma_{xB}\sigma_{zC} + \sqrt{2}\big[ \cos\gamma -
\cos(\alpha+\beta+\gamma) \big] \sigma_{yB}\sigma_{zC}
 -\big[ \sin(\alpha+\gamma)+\sin(\beta+\gamma) \big]
\sigma_{xB}\sigma_{xC} 
\qquad$$ $$
 +\big[ \cos(\alpha+\gamma)-\cos(\beta+\gamma) \big]
\sigma_{xB}\sigma_{yC} 
+\big[ \cos(\alpha+\gamma)+\cos(\beta+\gamma) \big]
\sigma_{yB}\sigma_{xC}
+\big[ \sin(\alpha+\gamma)-\sin(\beta+\gamma) \big]
\sigma_{yB}\sigma_{yC}
\Big\} \, . \; \eqno(48)$$

\noindent Note that (27) corresponds to the parameter choice
$\alpha=\beta=\gamma=0$. The Hamiltonian (48) describes
the three-spin XOR. For arbitrary values of $\alpha, \beta, \gamma$, the
interactions involved are all two-spin as desired.
The result, however, is not symmetric in any obvious way. It seems to
correspond to complicated tensor interactions involving expressions
of order two in the components of the three spins involved. No
rotational or other symmetry in the three-component spin space, or
planar symmetry, or uniaxial coupling, are apparent. All these would
correspond to the familiar Heisenberg, $XY$, and Ising couplings in
condensed matter physics. 

Thus, in order to realize interaction (48) in materials, a rather
anisotropic medium with highly nonsymmetric tensorial magnetic
interactions will be required. In this respect our analytical attempt
to design a multi-spin quantum gate cannot be considered really
successful. While we learned several general principles, the unappealing
form of the Hamiltonians (21) and (48) suggests that
different roots to the derivation of Hamiltonians should be also
explored. One could start with the more conventional magnetic
interactions, isotropic (Heisenberg), planar ($XY$), uniaxial (Ising),
and explore general-parameter Hamiltonians, adjusting the
coupling parameters numerically in search of those values for which
useful Boolean gate operations are carried out. Another approach
would be to use, in input and output, quantum states more general 
than the logic products of the qubit-basis-states which are presently
favored because of the correspondence with classical computers.
Then the design of Hamiltonians may be easier but there is a trade-off.
These devices will have to be incorporated into larger computers or
interact with classical (i.e., dissipative) computer components. 
Radically new programming ideas will be needed to use 
transformations that connect
{\it superpositions\/} of qubit states to carry out computations.

\acknowledgements

The work at Clarkson University has been supported in part by US Air
Force grants, contract numbers F30602-96-1-0276 and F30602-97-2-0089. 
The work at Rome Laboratory
has been supported by the AFOSR Exploratory Research Program and by
the Photonics in-house Research Program. This financial assistance 
is gratefully acknowledged.

{}

\vfill
\vfill
\hrule
\
 
\noindent\bf This article will be published in the Proceedings of the
Conference ``Photonic Quantum Computing. AeroSense 97'' (SPIE---The
International Society for Optical Engineering, 1997), SPIE Proceedings
Volume number 3076.

\end{document}